\documentclass{aa}
\usepackage{natbib}
\usepackage{graphicx}
\usepackage{times}
\usepackage{amsmath}
\usepackage{txfonts}
\usepackage{color}
\usepackage{ulem}
\renewcommand{\textbf}{\textrm}

\newcommand{\etal}{et al.,~}
\newcommand{\msun}{{\,\rm M}_{\odot }}
\newcommand{\lsun}{{\,\rm L}_{\odot}}

\newcommand{\hh}{\ifmmode {{\rm H}_2} \else {H$_2$} \fi}
\newcommand{\nht}{\ifmmode {{\rm NH}_3} \else {NH{\bas 3}} \fi}
\newcommand{\tco}{\ifmmode {^{13}{\rm CO}} \else {$^{13}{\rm CO}$}\fi}
\newcommand{\dco}{\ifmmode {^{12}{\rm CO}} \else {$^{12}{\rm CO}$}\fi}
\newcommand{\cdo}{\ifmmode {{\rm C}^{18}{\rm O}} \else {${\rm C}^{18}{\rm O}$}\fi}

\newcommand{\htco}{\ifmmode {{\rm H}^{13}{\rm CO}^{+} } \else {${\rm H}^{13}
{\rm CO}^{+}$ }\fi}
\newcommand{\hco}{\ifmmode {{\rm H}^{12}{\rm CO}^{+} } \else {${\rm H}^{12}
{\rm CO}^{+}$ }\fi}
\newcommand{\juz}{\ifmmode {{\rm J}=1\rightarrow 0} \else
{J=1$\rightarrow$0}\fi}
\newcommand{\jdu}{\ifmmode {{\rm J}=2\rightarrow 1} \else
{J=2$\rightarrow$1}\fi}
\newcommand{\jtd}{\ifmmode {{\rm J}=3\rightarrow 2} \else
{J=3$\rightarrow$2} \fi}
\newcommand{\jcq}{\ifmmode {{\rm J}=5\!\rightarrow\!4} \else
{${\rm J}=5\!\rightarrow\!4$} \fi}
\newcommand{\as}{\ifmmode {^{\scriptscriptstyle\prime\prime}}
\else $^{\scriptscriptstyle\prime\prime}$\fi}
\newcommand{\am}{\ifmmode {^{\scriptscriptstyle\prime}}
\else $^{\scriptscriptstyle\prime}$\fi}
\renewcommand{\hco}{\ifmmode {{\rm HCO}^+} \else {HCO$^+$} \fi}

\newcommand{\CI}{C{\small I}}

\newcommand{\unzero}{$^{3}P_1 \rightarrow {^{3}P_0}$}
\newcommand{\deuxun}{$^{3}P_2 \rightarrow {^{3}P_1}$}

\newcommand{\tabcqtau}{
\begin{table*}
\caption{Properties of CQ\,Tau, also named HD 36910.}
\label{tab:cqtau}
\begin{tabular}{lllllllll}
\hline
\hline
Spec.~type & T$_{eff}$\,(K) & Stellar lum.\,($\lsun$) & Distance$^{(*)}$\,(pc) & M$_*$$^{(**)}$\,($\msun$) & V$_{lsr}$$^{(**)}$\,(km/s) & PA ($^\circ$)$^{(**)}$& i ($^\circ$)$^{(**)}$& R$_{out}$(AU)$^{(**)}$\\
\hline
A8--F2 & 7200 & 8--16 & 140 & 1.8 & $6.2 \pm 0.04$ & $-36.7 \pm 0.3$&$29.3 \pm 1.7 $ & $210 \pm 20$\\
\hline
\end{tabular}\\
$^{(*)}$ assumed to be the Taurus distance.
$^{(**)}$ derived according to interferometric \dco\,J=2-1 data (Paper I)\\
\end{table*}
}

\newcommand{\tabcoor}{
\begin{table}
\caption{Sources coordinates (J2000).}
\label{tab:coord}
\centering
\begin{tabular}{lcc}
\hline
\hline
source & RA & DEC \\
\hline
CQ\,Tau & 05:35:58.485 & 24:44:54.19 \\
Mars$^{(*)}$ & 05:57:24 & 23:30:41 \\
Orion-Bar & 05:35:15.09 & -05:26:36.6 \\
\hline
\end{tabular}\\
$^{(*)}$ Aug.,25$^{th}$ 2009 at 0h00 (UT).
\end{table}
}

\newcommand{\tabresults}{
\begin{table}
\caption{Observed and predicted line flux}
\label{tab:flux}
\begin{tabular}{l|rr|rr|cc}
\hline
\hline
& \multicolumn{2}{c|}{\CI\ \unzero\ } & \multicolumn{2}{c|}{\CI\ \deuxun\ } & $\Sigma_{100}$ & p\\
& \multicolumn{2}{c|}{(Jy.km/s)} & \multicolumn{2}{c|}{(Jy.km/s)} & \multicolumn{2}{l}{($10^{17}$ cm$^{-2}$)} \\
\hline
Obs. & \multicolumn{2}{c|}{$< 6.6$} & \multicolumn{2}{c|}{$<25 $} & & \\
\hline
T$_k$ & 50\,K & 100\,K & 50\,K & 100\,K & \\
Model A & 30 & 48 & 68 & 135 & 12.0 & 2.3 \\
Model B & 13 & 15 & 33 & 52 & 2.3 & 1.5 \\
Model C & 26 & 37 & 60 & 111 & 8.7 & 1.4 \\
Thick & 60 & 128 & 125 & 300 & 1000 & 1.5 \\
\hline
Model J / B4 & \multicolumn{2}{c|}{8 -- 11} & \multicolumn{2}{c|}{24 -- 29} & & \\
\hline
\end{tabular}\\
Model A: $\chi=10^4$, g/d=100; B: $\chi=10^2$, g/d=10; C: Kurucz A3, g/d=10; Thick: flux for optically thick lines.
Predictions are for an assumed local line width of 0.2 km/s.
Model J is from \citet{Jonkheid_etal2007}, their model B4. 
All the line fluxes are scaled for a source distance of 140\,pc.
$\Sigma_{100}$ and $p$ are the results of the power law fitting of the modeled \CI\ surface densities (Fig. \ref{fig:coldens}).
\end{table}
}

\newcommand{\figcoldensok}{
\begin{figure*}
\centering
\includegraphics[width=4.7cm,angle=270]{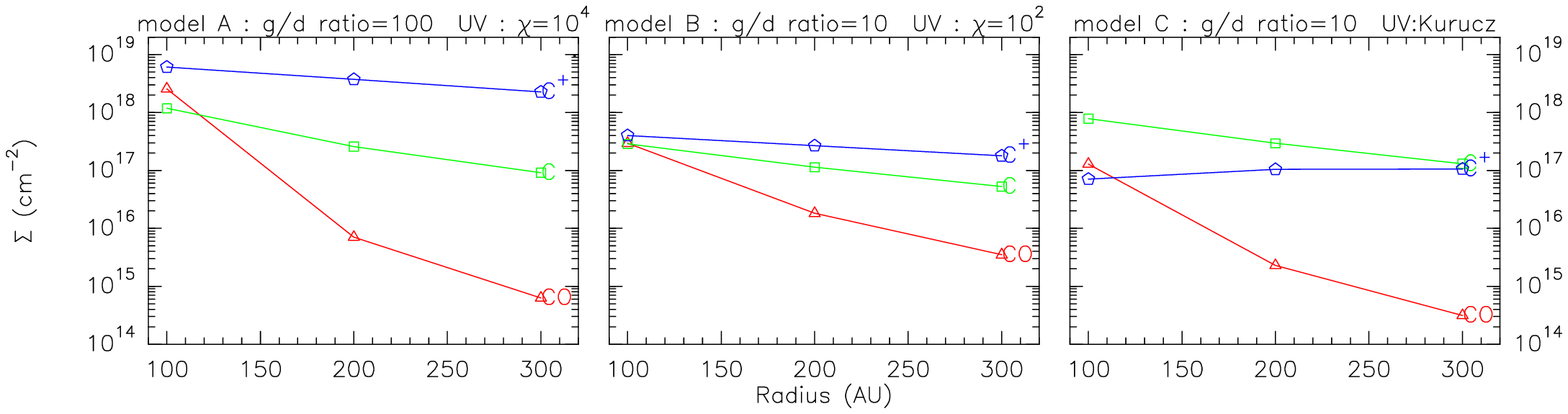}
\caption{
Radial distribution of the surface density of C$^+$, C, and CO for three models which are in agreement with the CO observations
}
\label{fig:coldens}
\end{figure*}
}

\newcommand{\figcoldensbad}{
\begin{figure*}
\centering
\includegraphics[width=4.7cm,angle=270]{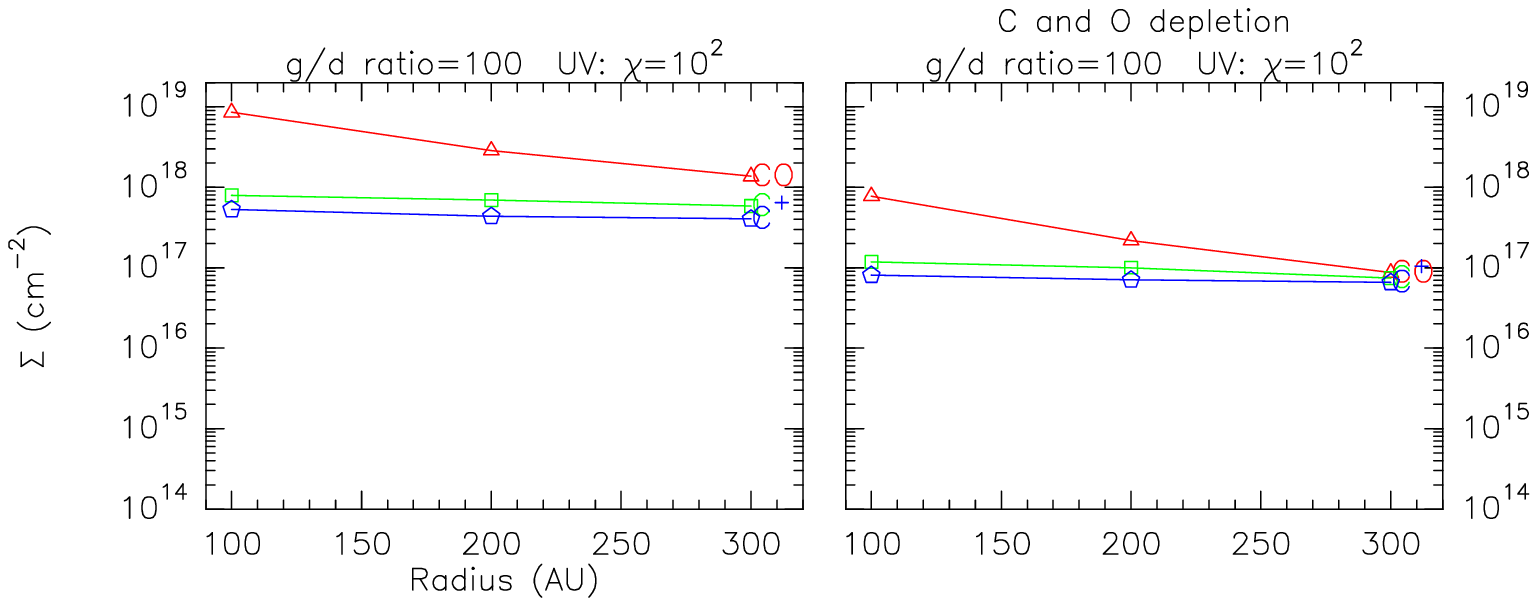}
\caption{
Radial distribution of the surface density of C$^+$, C, and CO for two models which are NOT in agreement with the CO observations
}
\label{fig:coldensbad}
\end{figure*}
}

\newcommand{\figabondance}{
\begin{figure*}
\centering
\includegraphics[angle=270,width=18cm]{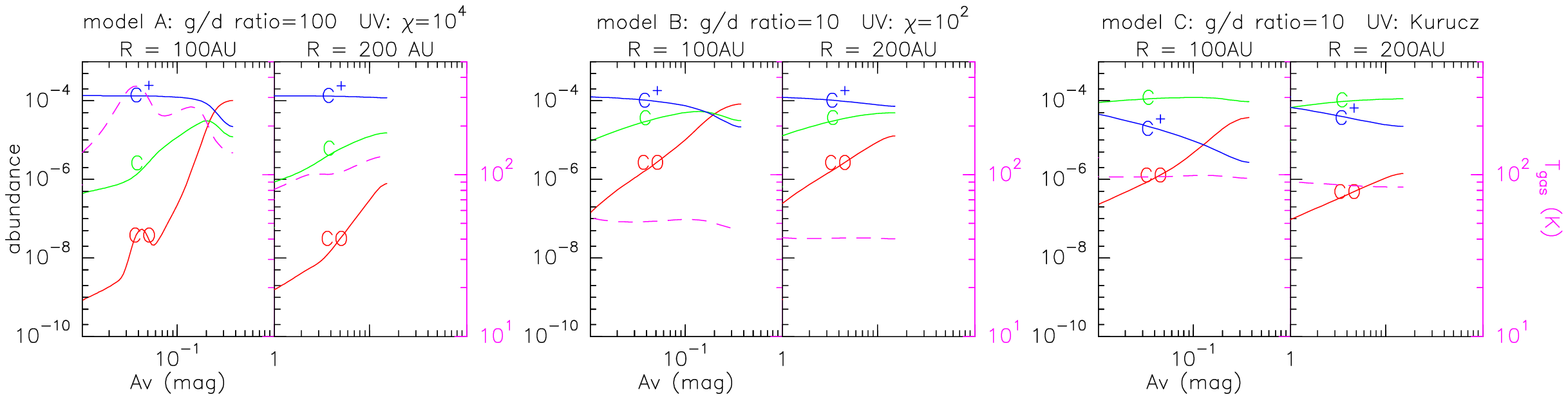}
\caption{Vertical distribution of the abundance of C$^+$ (blue), C (green), and CO (red) and gas temperature (pink dashed lines) for the three models which are in agreement with the CO observations.}
\label{fig:abondance}
\end{figure*}
}

\newcommand{\oldtabchimie}{
\begin{table}
\caption{Elemental abundances with respect to total hydrogen}
\centering
\begin{tabular}{lc}
\hline
\hline
Element & abundance\\
\hline
\vspace{-0.25cm}\\
C & $1.38 \times 10^{-4}$\\
O & $3.02 \times 10^{-4}$\\
N &$7.95 \times 10^{-5}$\\
Mg &$1.00 \times 10^{-8}$\\
S &$2.00 \times 10^{-6}$\\
Si &$1.73 \times 10^{-8}$\\
Fe &$1.70 \times 10^{-9}$\\
\hline
\end{tabular}
\label{tab:abon-chimi}
\end{table}
}

\newcommand{\plotorion}{
\begin{figure}[t!]
\centering
\includegraphics[height=8.25cm,angle=270]{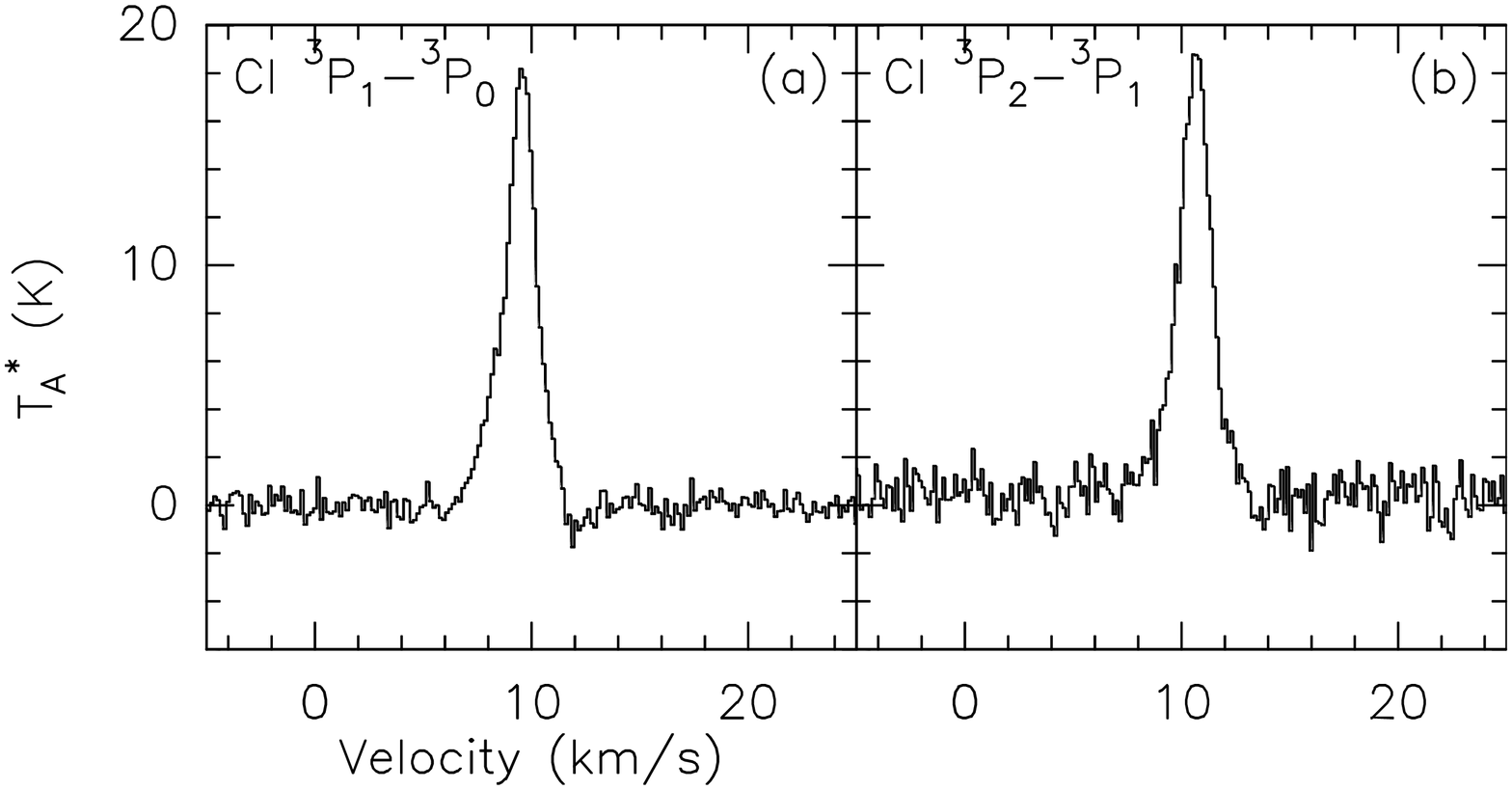}
\caption{\CI\ $^3P_1$$\rightarrow$$^3P_0$ (a) and $^3P_2$$\rightarrow$$^3P_1$ (b) spectra towards the Orion Bar
}
\label{fig:orion}
\end{figure}
}

\newcommand{\plotcqtau}{
\begin{figure}[t!]
\centering
\includegraphics[height=8.25cm,angle=270]{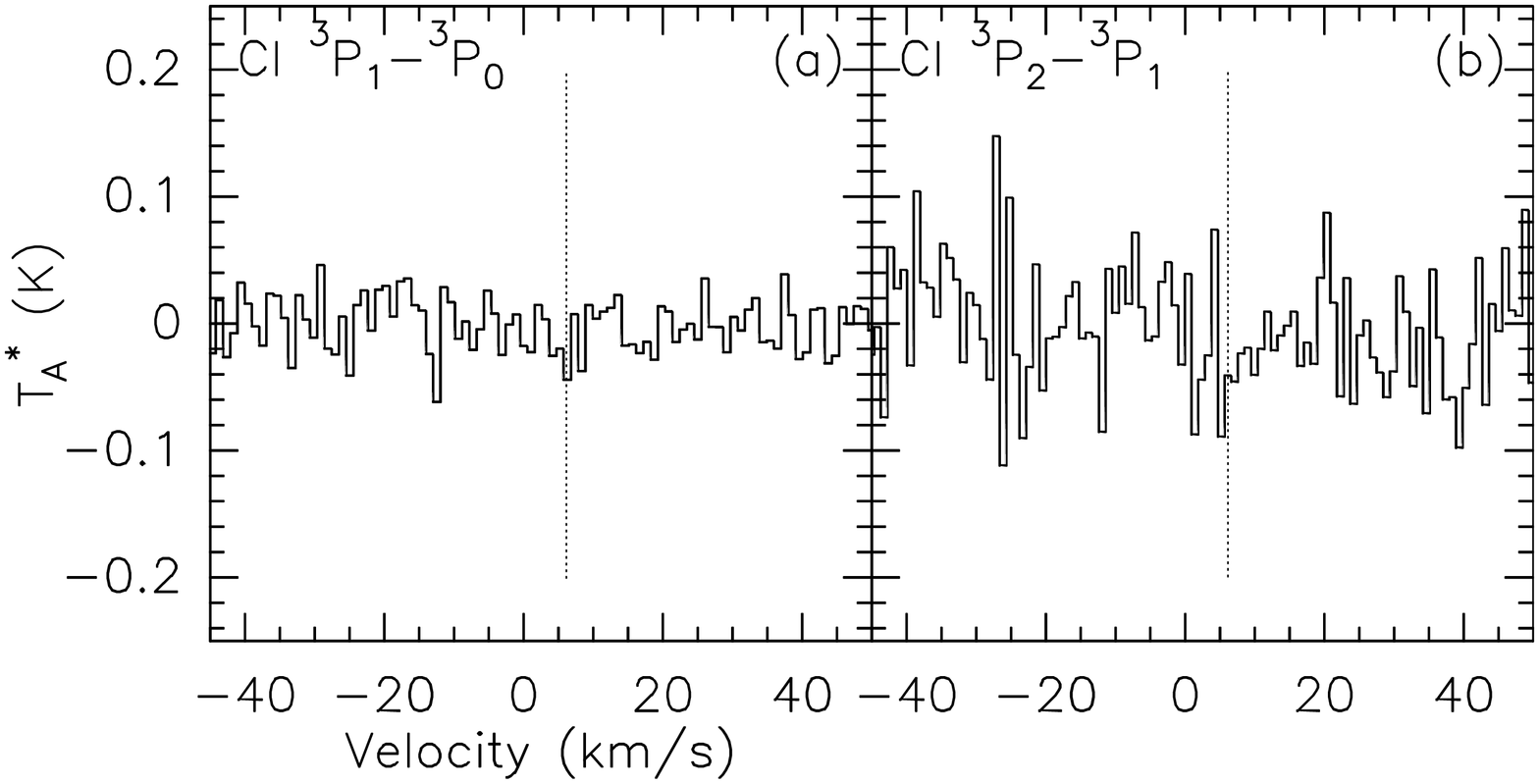}
\caption{(a) \CI\ $^3P_1$$\rightarrow$$^3P_0$ spectra, rms=0.020\,K and (b) \CI\ $^3P_2$$\rightarrow$$^3P_1$ spectra, rms=0.046\,K towards CQ\,Tau. 
 The spectral resolution is about 1\,km/s. The dotted line correspond to the V$_{lsr}$ of the source.}
\label{fig:spectre}
\end{figure}
}

\begin{document}

\title{C\,I observations in the CQ\,Tau proto-planetary disk: \\ evidence for a very low gas-to-dust ratio ?
\thanks{Based on observations carried out with the Atacama Pathfinder Experiment. APEX is a collaboration between the Max-Planck-Institut f\"ur Radioastronomie, the European Southern Observatory, and the Onsala Space Observatory.}}
%
\author{
Edwige Chapillon\inst{1}
\and B\'ereng\`ere Parise\inst{1}
\and St\'ephane Guilloteau\inst{2,3}
\and Anne Dutrey\inst{2,3}
\and Valentine Wakelam\inst{2,3}
}
\institute{
MPIfR, Auf dem H\"ugel 69, 53121 Bonn, Germany\\
\email{echapill@mpifr-bonn.mpg.de, bparise@mpifr-bonn.mpg.de}
\and
Universit\'e de Bordeaux, Observatoire Aquitain des Sciences de l'Univers, BP 89, F-33271 Floirac Cedex, France
\and CNRS, UMR 5804, Laboratoire d'Astrophysique de Bordeaux, BP 89, F-33271 Floirac Cedex, France\\
\email{guilloteau@obs.u-bordeaux1.fr, dutrey@obs.u-bordeaux1.fr, wakelam@obs.u-bordeaux1.fr}
}
\offprints{E.Chapillon \email{echapill@mpifr-bonn.mpg.de}}
\date{Received xx-xxx-xxxx, Accepted xx-xxx-xxxx}
\authorrunning{Chapillon \etal\ }
\titlerunning{\CI\ in CQ\,Tau}

\abstract
{Gas and dust dissipation processes of proto-planetary disks are hardly known.
Transition disks between Class II (proto-planetary disks) and Class III (debris disks) remain difficult to detect.}
{We investigate the carbon chemistry of the peculiar CQ\,Tau gas disk. It is likely a transition disk because it exhibits weak CO emission with a relatively strong millimeter continuum, indicating that the disk might be currently dissipating its gas content.}
{We used APEX to observe the two \CI\ lines \unzero\ at 492\,GHz and \deuxun\ at 809\,GHz in the disk orbiting CQ\,Tau.
We compare the observations to several chemical model predictions. We focus our study on the influence of the
stellar UV radiation shape and gas-to-dust ratio.}
{We did not detect the \CI\ lines. However, our upper limits are deep enough to exclude high-\CI\ models. The only available models compatible with our limits imply very low gas-to-dust ratio,
of the order of a few, only.}
{These observations strengthen the hypothesis that CQ\,Tau is likely a transition disk and suggest that gas disappears before dust.}


\maketitle{}


\section{Introduction}

Gas and dust disks orbiting Pre-Main-Sequence stars are believed to be the birth-place of planetary systems.
Several disks surrounding young stars have been detected and studied both in continuum and molecular lines at mm/submm
wavelengths \citep[e.g.][and references therein]{Pietu_etal2007}.
However, little is known about how their gas and dust content dissipate.
Transition disks between Class II (proto-planetary disks) and Class III (debris disks) remain hard to detect.
Accretion, viscous spreading, planet formation, and photo-evaporation (enhanced by grain growth and dust settling) should play
a role in the disappearance of the gas and dust disk, but their relative importance has not been estimated yet.

Disks with strong mm continuum excess and weak CO lines are interesting in this respect, as they may represent a stage in which gas is being dissipated, while large dust grains responsible for the mm continuum still remain, perhaps settled in the mid-plane. The first case discovered was BP\,Tau which exhibits a warm (50 K), small (radius $\sim$ 120\,AU) disk with a very low CO to dust ratio \citep{Dutrey_etal2003}, corresponding to an apparent depletion factor 100 for CO. \citet[][hereafter Paper\,I]{Chapillon_etal2008} have recently found that the Herbig\,Ae (HAe) stars CQ\,Tau and MWC\,758 have disks with similar characteristics.

\citet{Chapillon_etal2008} showed that such a low CO to dust ratio can be explained by photo-dissociation by UV photons coming from the star with a standard gas-to-dust mass ratio. However, solutions with low gas-to-dust ratio also exist. These solutions differ by the amount of atomic and/or ionized carbon which result from the CO photo-dissociation. Thus, detection of either \CI\ or C$^+$ provides a useful diagnostic on the disk structure. 
An upper limit on the \CI\ \deuxun transition in the Herbig Be star HD\,100546 has been recently reported by \citet{Panic+etal_2010}.

We report here a search for \CI\ in CQ\,Tau. We present our observations in Section 2. Our results are discussed and compared with chemical modeling in Section 3. We summarize in the last section.

\section{Observations and results}

We selected CQ\,Tau as our main target, because it appears to be the warmest source: the mm continuum and $^{12}$CO observations of Paper\,I indicate a gas temperature $> 50$~K, a disk outer radius near 250\,AU, and an apparent CO depletion of 100.
CQ\,Tau (see Table 1) is one of the oldest HAe star ($\sim$ 6 -- 10\,Myr, although a revision of its distance
implies that CQ\,Tau is younger than initially thought, see Paper\,I) surrounded by a resolved disk \citep{Mannings_Sargent1997, Doucet_etal2006}.
This star exhibits an UX-Ori-like variability \citep{Natta_etal1997}, in contradiction with the observed low disk inclination (Paper\,I). Its dust emissivity exponent is about 0.7 \citep[Paper\,I and][]{Testi_etal2003}, indicating significant grain growth.
\tabcqtau


We observed simultaneously the \CI\ $^3P_1$$\rightarrow$$^3P_0$ and $^3P_2$$\rightarrow$$^3P_1$ lines in CQ\,Tau with FLASH \citep{Heyminck_etal2006} on  APEX during August 2009 under excellent weather conditions (pwv $\leq 0.3$mm). We used wobbler mode (60$''$). The total integration time on source is 67\,min.
Pointing and focus were frequently checked on Mars (the planet was very close to the source, see Table \ref{tab:coord}). 
Receiver tuning was checked on the Orion-Bar (Fig.\,\ref{fig:orion}).
Our observation of \mbox{\CI\,$^3P_1 $$\rightarrow$$^3P_0$ } towards the Orion Bar peaks at T$_\mathrm{A}^*\simeq 17.5 K$. Taking the beam efficiency measured toward the Moon $\sim 0.8$ (the emission being extended) this leads to a peak temperature of T$_\mathrm{MB} \sim$ 21\,K.
This is in quite good agreement with previous observations of \CI\ at 492\,GHz toward the Orion Bar by \citet{Tauber_etal1995} with the CSO 10.4\,m dish (15$''$ beam, our pointing is between their ``d'' and ``e'' spectra, i.e.  $ 17 \leq T_\mathrm{MB} \leq 25$\,K, see their Figs.\,1 and 2).

The spectra towards CQ\,Tau are presented in Fig.\,\ref{fig:spectre}. We did not detect any \CI\ emission. 
 The 1$\sigma$  sensitivity on the integrated area is given by the following formula
\mbox{$  \sigma \textrm{(Jy.km.s$^{-1}$)} = S \textrm{(Jy/K)}   \sigma \textrm{(K)}  \sqrt{ \delta \mathrm{v} \Delta \mathrm{v} }$}
 with $\delta\,\mathrm{v}$ the spectral resolution and $\Delta\,\mathrm{v}$ the line width. We take \mbox{$\Delta\,\mathrm{v}\,=\,5$\,km.s$^{-1}$} according to the PdBI CO observations (Paper\,I). The rms noise in Fig.\,\ref{fig:spectre} are (1$\sigma$) rms 0.020 and 0.046\,K (T$_\mathrm{A}$) for a spectral resolution of 1.04 and 0.95 km.s$^{-1}$ at 492 and 809 GHz, respectively. Taking the conversion factors between Jansky and Kelvin equal to 48 and 70\,Jy/K, we thus obtain $3\sigma$ upper limits of 6.6\,Jy.km.s$^{-1}$ and 25\,Jy.km.s$^{-1}$ for the $^3P_1$$\rightarrow$$^3P_0$ and $^3P_2$$\rightarrow$$^3P_1$ lines (see Table\,\ref{tab:flux}).

\plotorion
\plotcqtau

\tabcoor

\section{Discussion}

\subsection{Chemical Modeling}
\oldtabchimie
\figcoldensok
\figabondance


We use the PDR code from the Meudon group \citep{leBourlot_etal1993,lePetit_etal2006} with the modification of the extinction curve calculation according to the grain size distribution described in Paper I.  The chemical network and elemental abundances (given in Table \ref{tab:abon-chimi}) are similar to that of \citet{Goicoechea_etal2006}. No freeze-out onto grains is considered, as the gas temperature determined by CO observation is $>50\,K$. We solve the radiative transfer perpendicular to the disk plane for each radii, with an incident UV flux scaling as $1/r^2$. Radiative transfer, chemistry and thermal balance are consistently calculated. For more details and a description of the use of the code see Paper\,I.

From Paper\,I, our best models to reproduce the CO depletion and gas temperature in CQ\,Tau are 1) a disk illuminated by a strong UV field from the star (Draine field with a scaling factor $\chi\,=\,10^4$ at R\,=\,100\,AU) with a normal gas-to-dust ratio (g/d\,=\,100, model A) or 2) a disk illuminated by a weaker UV field ($\chi\,=\,10^2$) with a modified g/d ratio of 10 (model B). In both cases we considered big grains with a maximum size of 1\,cm as suggested by the previous dust and CO observations. The computed surface densities and abundances of CO, \CI\ and C{\small II} are shown in Figs.\,\ref{fig:coldens} and \ref{fig:abondance} left and middle panels. 
A cut perpendicular to the disk mid-plane in our models A and B shows three parts (Fig.\ref{fig:abondance}).  At 100\,AU radius, C$^+$ is the most abundance form of carbon in the upper layer, then there is a layer transition where \CI\ is quasi-dominant and finally CO is the main carbon-component in the mid-plane. At larger radii (200\,AU), C$^+$ stays however the dominant carbon-component even in the disk mid-plane, because the total opacity perpendicular to the disk mid-plane is low.\\

\subsection{The UV problem}
As indicated in Paper\,I (\S\,5.2), the actual UV spectra of HAe stars are not well known.
Observations of MWC\,758 indicate a UV spectrum shortwards of 1500\,$\AA$ whose shape is better represented by a Draine field than by a purely photospheric spectrum (Paper\,I). From spectra obtained by the IUE satellite, the CQ\,Tau UV flux is 10--30 times lower than that of MWC\,758 \citep{Grady_etal2005,Blondel_djin2006}, but the shape is unknown. To check the importance of the UV spectrum profile, we performed another model assuming a photospheric spectrum from a A3 type star (Kurucz database\footnote{http://kurucz.harvard.edu/}).
CQ\,Tau is a somewhat cooler star (A8 to F2), so this is most likely an overestimate of the UV flux.

The CO abundance appears very sensitive to the UV profile. In absence of UV excess and in the case g/d=100 the chemistry is dramatically affected. The disk is mainly molecular and the C/CO transition occurs at low opacities (i.e. closer to the disk atmosphere). This model leads however to too high CO surface densities (see Fig. \ref{fig:coldensbad}, left panel). With g/d=10 (Model C, see Figs.\ref{fig:coldens} -- \ref{fig:abondance}), the CO surface densities are of the same order of magnitude as in model B and therefore also compatible with CO observations frome Paper\,I, but now the gas is warmer and there is more \CI\ than C$^+$ (\CI\ being the main carbon form in the mid-plane).

\citet{Bergin_etal2003} have shown that the UV excess of T\,Tauri stars is dominated by strong lines emission, in particular \mbox{Lyman\,$\alpha$}.
We therefore explored the influence of the shape of the UV spectrum by studying the impact of the  \mbox{Lyman\,$\alpha$} line at 1215\,$\AA$. We performed a couple of runs with a modified stellar spectrum, i.e. a photospheric spectrum with an additional line, and the two gas-to-dust ratios (10 and 100). The effects on the surface densities of CO, C and C$^+$ were found to be negligible.

\subsection{Elemental depletion}

Another possibility to explain the lack of CO and \CI\ in this object is that, even with a normal g/d, the C and O elemental abundances may be lower than the ones adopted here (e.g. by depletion onto grains surfaces). We made a model with a stellar UV field including UV excess ($\chi=10^2$), a standard value of g/d (100) and an elemental depletion factor of 10 of heavy elements (see Fig. \ref{fig:coldensbad}, right panel). The amount of \CI\ is in reasonable agreement with our observation but the CO is largely overestimated. The larger amount of CO compared to the g/d=10 and standard abundances case is due to mutual shielding by H$_2$. 
An unlikely elemental depletion of $\sim\,100$ is required to match the observations of both CO and \CI.

\figcoldensbad

\tabresults

\subsection{Spectrum Predictions}

 Characteristic flux densities can be recovered by noting that an approximate integrated line flux is given by
$$S = B_\nu(T_{ex},\nu) \pi \left(R_{out}/{D}\right)^2 \rho \delta V \mathrm{cos}\,i$$
where $B_\nu$ is the Planck function, $T_{ex}$ is the excitation temperature, $R_{out}$ is the outer radius, $D$ is the distance of the star, $\delta V$ is the local (i.e. sum of thermal + turbulent) line width.
For optically thin lines $\rho \simeq \tau$, the opacity, and for optically thick lines $\rho$ is of the order of a few \citep[see][their Eq.2 and Fig.4]{Guilloteau_Dutrey1998}.
In the optically thick case, $\rho \delta V \leq 2 V_\mathrm{out}$, where $V_\mathrm{out}$ is the Keplerian velocity at the disk outer radius, about 5.5 km.s$^{-1}$ in our case. In the optically thin regime, as $\tau \propto \Sigma/\delta V$, where
$\Sigma$ is the surface density, the emission is just proportional to the number of molecules.

We used the radiative transfer code DISKFIT optimized for disks \citep{Pietu_etal2007,Pavlyuchenkov_etal2007}, to make more accurate predictions for the \CI\ line profiles and integrated intensities from our models. The model surface densities were extrapolated using simple power laws
  \mbox{$\Sigma(r) = \Sigma_{100} (r/100\mathrm{\,AU})^{-p}$}. The disk parameters (size, temperature, Keplerian velocity, inclination, turbulent width) were taken from Paper\,I. Our values of $\Sigma_{100}$ and $p$ and the line flux predictions using $T_k\,=\,50$ and $100$\,K are given in Table\,\ref{tab:flux}, with the observed upper limits. $T_k\,=\,100$~K is obtained from the best fit model of  the J=2-1 CO line. 50\,K is a lower limit, but this temperature provides a better fit to the CO J=3-2 spectrum obtained at JCMT by \citet{Dent_etal2005}. Non LTE effects are expected to be negligible, because the \CI\ transitions have low critical densities of the order of $<\,10^4$\,cm$^{-3}$ \citep{Monteiro_Flower1987, Schroder_etal1991}, while in our model, the \CI\ layer is located in a high density region ($\sim\,10^5$ -- $10^6$\,cm$^{-3}$).

Considering the uncertainty in the temperature, the non detection of both \CI\ lines excludes the high \CI\ models (A and C). It is just marginally consistent with the low g/d, non photospheric UV spectrum model (B), provided the kinetic temperature does not exceed 50\,K. In this case, the \CI\ lines are mostly optically thin, and only the line wings which come from the inner regions because of the Keplerian rotation are optically thick.

\subsection{Comparison with other models}
Although there is no specific modeling of the CQ\,Tau disk and star system, it is worth comparing our results with predictions from other chemical studies.

The chemistry of HAe disks has also been studied by \citet{Jonkheid_etal2007}, with different assumptions than in our study.
 The UV transfer is performed with a 2D code \citep{vanZadelhoff_etal2003} whereas in our case this is a 1D method. The self-shielding of H$_2$ and CO are calculated assuming a constant molecular abundance whereas we compute it explicitly. Moreover, they mimic the dust growth (and/or settling) decreasing the mass of the small grains (i.e., interstellar grains) and they consider PAHs. The radiation field is a NEXTGEN spectrum (which is a pure photospheric spectrum). Low CO amounts can be reproduced by their model B4.
 They predict that \CI\ is the main form of carbon (their Fig. 6), in agreement with our model C (photospheric spectrum and low gas-to-dust ratio). Jonkheid \etal model B4 yields integrated line intensities of about 1 -- 1.3 K.km.s$^{-1}$ for a beam size of 6.7$''$, an inclination of 45$^\circ$ and a turbulent width of 0.2 km.s$^{-1}$. This corresponds to 8 -- 11 and 24 -- 29 Jy.km.s$^{-1}$ for the 492 and 809 GHz lines, respectively (Table \ref{tab:flux}).
Taking the 1.3\,mm flux from CQ\,Tau and using an absorption coefficient of 2 cm$^2$.g$^{-1}$ (per gram of dust) with a dust temperature of 50 K, the estimated dust mass is $4.6\,10^{-5} \msun$. This is likely a lower limit, as the adopted absorption coefficient is large. The \textit{gas mass} in the most appropriate model from \citet{Jonkheid_etal2007} is $10^{-4}\msun$, which would imply a gas-to-dust ratio of $\sim 2$, only. Our best model (B) with g/d=10 predicts somewhat too strong lines. From Table \ref{tab:flux}, as the \CI\ lines are close to the optically thin regime, a reduction of g/d by a factor of 2-3 may bring model and observations to agreement.
Thus, whatever chemical model of HAe disk we consider, the CO observations and the measured upper limits on both \CI\ lines can only be reproduced by a very low gas-to-dust ratio. 

Although a factor of a few may be within the modeling uncertainties, it appears difficult to reconcile the current constraints on CO and \CI\ with gas to dust ratio larger than 10.\\

The influence of X-Rays is not taken into account in our PDR code, neither in \citet{Jonkheid_etal2007}. Although their $L_X/L_\mathrm{bol}$ are lower, on average, HAe stars are as strong X-ray emitters as T\,Tauri stars \citep{Skinner_etal2004}. No specific model exist for HAe disks, but studies appropriate to T\,Tauri stars indicate that X Rays are expected to increase the \CI\ intensity \citep{Meijerink_etal2008,Gorti_Hollenbach2008,Ercolano_etal2009}. We thus do not expect the (unknown) X-ray luminosity of CQ\,Tau to affect our conclusions.

Another alternative would be that models over-predict \CI~ due to inaccurate treatment of the physical processes. 
While this cannot be excluded, we note however that the two existing HAe disk chemical models make very different approximations (one on the radiative transfer, the other on the self-shielding), and yet converge towards similar predictions.

\section{Summary}
Our observations of the \CI\,$^3P_1$$\rightarrow$$^3P_0$  and $^3P_2$$\rightarrow$$^3P_1$ transitions in the CQ\,Tau disk leads to significant upper limits which allow us to reject chemical models producing a high amount of \CI. The absence of \CI\ emission together with the low CO emission can be explained by currently available chemical models of HAe disks, only if the gas-to-dust ratio is notably low (of order 2-5) in the CQ\,Tau disk. This suggests that the CQ\,Tau disk is in a transition stage between the proto-planetary and debris disk phases.
This work also allows us to conclude:
\begin{enumerate}\itemsep 0pt
\item PDR models which match the CO data are qualitatively compatible with our upper limits, but still overestimate the total amount of C\,I unless the gas-to-dust ratio is of the order of a few.
\item The \CI\ surface density is sensitive to the shape of the UV spectrum, with neutral carbon being the dominant species in the absence of UV excess.
\item Introducing the Lyman $\alpha$ line in the stellar UV spectrum does not change the \CI\ surface density.
\item All models matching the low CO and \CI\ intensities predict that Carbon should be mainly in the ionized form C$^+$.
That could be tested by Sofia or Herschel observations.
With a complete census of CO, \CI\, and C$^+$, the overall surface density of gas-phase carbon will be constrained, allowing a better estimate of the g/d ratio.
\end{enumerate}
While the above conclusions are given under the assumption that the chemical models of HAe disks are indeed valid, observing more disks, especially large disks with bright CO, would provide a useful test of their validity.

\begin{acknowledgements}
E.C. and B.P. are supported by the {\it Deutsche Forschungsgemeinschaft} (DFG) under the Emmy Noether project PA 1692/1-1.
S.G., A.D. and V.W. are financially supported by the French program ``Physique Chimie du Milieu Interstellaire'' (PCMI) from CNRS/INSU.
\end{acknowledgements}

\bibliography{bib}
\bibliographystyle{aa}

\end{document}